\documentstyle[twocolumn,aps]{revtex}
\begin{document}
\draft
\title{Atom Interferometers: Beyond  Complementarity 
Principles }
\author{ Zhi-Yuan Li } 
\address{Ames Laboratory, US Department of Energy and
Department of Physics and Astronomy, Iowa State University, 
Ames, Iowa 50011, and
Institute of Physics, Chinese Academy of Sciences, P. O. Box 603,
Beijing 100080, China}
\date{September 4,  2001}
\maketitle

\begin{abstract}
  Complementarity lies at the heart of    
conceptual foundation of orthodox quantum mechanics. The wave-particle
duality  makes it impossible to tell 
which slit each particle passes through and still observe an 
interference pattern in a Young's double-slit experiment.  
In this paper,  this duality  
is appraised under the standard formulation of  quantum mechanics
for atom interferometers. It is found that
the internal freedoms like electronic states
of an atom can be used to  detect the which-path information of each
 atom while uphold the interference  pattern of atoms when  
 the center-of-mass motion of atoms is detected. 
 
\end{abstract}
\narrowtext
\par
The principle of complementarity represents one of the basic 
conceptual foundation of orthodox quantum mechanics and
its Copenhagen interpretation. One  classic example is
the wave-particle duality of matter, where  the position (particle-like)
and momentum (wave-like) are complementary attributes of a quantum
object, and precise knowledge of one of them  implies all possible outcomes
of the other one are equally probable. This wave-particle duality
is most clearly revealed in  Young's double-slit experiment. There
one finds that it is impossible to tell which slit each particle passes
through while still observe an interference pattern. In another word,
once the which-path information is possible,  immediately the interference
pattern disappears, and vice versa. 

\par
Historically, the wave-particle duality is enforced with the aid of
 the position-momentum uncertainty relation. 
This mechanism can be found in the thought experiments of
Einstein's recoiling slit [1], Feynman's  light-electron 
scattering scheme [2], and Heisenberg's $\gamma$-ray microscope [3], 
where the act of measurement on which-path information inevitably
disturbs the particle in such a manner  that 
coherence is lost and the interference pattern disappears.  However,
in recent years,  people  have found ways around the uncertainty 
principle using atom interferometers with the aid of
modern quantum-optical techniques [4-10], while still 
manifesting the complementarity principle through entanglement 
between the which-path detector and  the atomic motion. But some 
other authors
have argued that  loss of interference requires that there must be some 
classical or quantum momentum transfer in accordance 
with the uncertainty principle [11,12].  Note that 
 the controversy here is simply caused 
by the assumption (or hypothesis) that the 
complementarity principle  should  always keep true under any
situations, no matter how much the detector is improved upon
the original ones from which Bohr invented this  principle 
more than 70 years ago.  

\par
In this paper, we will analyze recent results on  
atom interferometers [4-10]
 in the framework of standard formulation of quantum mechanics. 
After a careful look at how the interference pattern
is formed in these experiments,
we will show that these atom interferometers have implied, contrary to
what these experiments are intended to prove,
the  possibility beyond the complementarity principle. Namely,
it is possible to tell which-path information and
observe the interference pattern simultaneously in atom interferometers.

\par
Before  going into details of our analysis, we ask one 
 question which we think is very important for clarifying
 the complementarity principle, 
``What on earth determines the interference pattern 
of  atoms formed under a 
certain experimental setup?" The natural answer to this question
is  the motion of the atom itself, or more particularly, 
the atomic  center-of-mass, which is the most appropriate
 representation of the atom itself as  a quantum object. 
As is well-known,
an atom always possesses some  internal structures and  freedoms. 
 For example, an atom consists of a nucleus at its center 
and some electrons around the nucleus. The nucleon
consists of  many protons and neutrons, which in turn are made up of quarks.  
When observed, the atom  might interact 
with the environment (e.g., which-path detectors)
   through some of its internal  freedoms during 
the process of motion. 
It is then obvious that when we talk about the interference of atoms,
  we should only refer to the center-of-mass itself,  we never refer to
the interference of any part of these internal composites. 
Otherwise, incoherence
in any internal freedom (such as quarks) will force one to conclude that
 the atomic wave can not be coherent, which is of 
course not the case. The effect of
internal freedoms is essentially reflected through their influence 
on the motion of the center-of-mass. The same principle  
applies to the environment. 

\par
The internal structures or freedoms of an atom, when
coupled  with  external detectors, provide an useful
information on the atomic motion, since they are tightly bound with the
atomic center-of-mass. In addition, the coupling  can be controlled
to be strong enough to cause significantly observable change to these freedoms, 
while  weak enough not to 
change the motion of the far more massive center-of-mass.  
This constitutes the first category of detecting the motion of an atom.
The most convenient among these freedoms in an atom  is electrons
at the most outer shell,
 who  have some different states,  and the transition between these states
caused by interactions with the environment only
 causes negligible influence to the center-of-mass motion. 
A notable example is  Rydberg atoms with very high 
excited electronic  states placed inside a micromaser cavity. The cavity
can serve as the which-path detector for both the atomic and electronic
states through recording variation of the photon field. The electronic state 
is another which-path  detector avaliable for the  atom. 
One can find that recent atom interferometers [4-10] exactly 
work within this category. 

\par
The second category to detect the atomic  motion is direct 
observation of the atomic center-of-mass. The classical examples [1-3] 
taken to demonstrate
the  complementarity principle essentially lies inside this category. 
Naturally, it always takes  much  stronger interaction 
to directly change the state of the
center-of-mass than to change the state 
of an electron much lighter in mass.
 Then it is much harder to control the atomic motion than the bound
electronic state. When this kind of detector  is utilized
to detect the which-path information of particles (atoms,
electrons, photons, etc.) 
in the double-slit experiments, the act of measurement  inevitably
disturbs the particle  in such an uncontrolled manner  
that  coherence of 
the incident wave is  lost and the interference pattern disappears.  
This mechanism can be well described qualitatively 
by the Heisenberg's position-momentum uncertain relation. Detection 
on atom requires localization in a far smaller region (thus far bigger 
momentum uncertainty) and causes much stronger disturb on the atomic state
than detection on electronic states does.
In regard to this category we agree with Feynman by 
 saying (upon his light-electron scheme):
If an apparatus is capable of determining which hole the 
electron goes through,
it cannot be so delicate that it does not disturb the pattern in
an essential way. No one has ever found  (or even thought of)
 a way around the uncertainty principle.

\par
Above qualitative discussion can be made  mathematically
more precise using the standard formulation of quantum mechanics.
Assume that the which-path detector is placed right before the
double-slit, and after passing the double-slit 
the atomic beam  propagates in free space before  arriving at the
interference-pattern detector.
 The interference pattern is determined by the
wave function right at the double slit, which,  under an free-space
evolution (satisfying Kirchhoff's law), yields
$$\Psi({\bf r}) \propto \psi_1({\bf r})+\psi_2({\bf r}), \eqno(1)$$
where $\psi_1$ ($\psi_2$)  is the wave function  
evolving from slit 1 (2).  The interference  pattern is calculated by 
$P({\bf r})=|\Psi({\bf r})|^2$. Eq. (1) applies to both
the atomic (center-of-mass) and  electronic motion, 
$$\Psi_c({\bf r}) 
\propto \psi_{1c}({\bf r})+\psi_{2c}({\bf r}), \eqno(2)$$ 
$$\Psi_e({\bf r}) \propto \psi_{1c}({\bf r})\psi_{1e}
+\psi_{2c}({\bf r }) \psi_{2e}. \eqno(3)$$
Here $\psi_{1c}$ ($\psi_{2c}$) and $\psi_{1e}$ ($\psi_{2e}$) are
the wave function of the atomic center-of-mass and the corresponding
electronic state in path 1 (2), respectively.
In deriving Eq. (3) we have used the Born-Oppenheimer approximation,
where the  electronic state is  adiabatically adapted to the
center-of-mass motion, namely, the electron is tightly bound with the
atomic center-of-mass. Eqs. (1)-(3) provides an unified framework to
discuss various  types of double-slit experiments. 

\par
First, the incident  atomic beam  is incoherent.  In this case we have 
$\Psi_c({\bf r}) \propto \psi_{1c}({\bf r})e^{i \delta_1}
+\psi_{2c}({\bf r})e^{i \delta_2},$ where $\delta_1$ and $\delta_2$
are two independent random phases, no matter whether or not 
the which-path detector is present. The interference pattern 
is simply $P({\bf r})=|\psi_1|^2+|\psi_2|^2$ after 
ensemble average, with the fringe  completely washed out.

\par
Second, we have a coherent  atomic beam incident on the double-slit
after passing through a which-path detector working with
 the atomic center-of-mass as in the classical examples [1-3]. 
Due to uncontrollable interaction, we can assume that
the mere result of measurement is  
to introduce two independent radom phases to the atomic center-of-mass
wave function at the two slits. 
Therefore, the interference pattern also  disappears.
 The atomic beam emerging from the double-slit now becomes incoherent.
According to Eq. (3), this incoherence in the center-of-mass 
motion in turn destroys   the electronic interference pattern.

\par
Third, a coherent  atomic beam is incident on the double-slit, and now
the which-path detector is working with the bound electronic states.
In the first case, the electron-detector interaction  is so
strong that the atomic center-of-mass wave function is severely 
disturbed. Then we also find an invisible interference pattern
formed by both the atoms and electrons. For example, the coherent 
atomic  beam  can be made incoherent by  spontaneous
emission and reabsorption of photons by many times. It is clear
that in order to obtain visible interference pattern, the 
electron-detector interaction  must be designed
such that the influence on the atomic motion is negligible. 
Several schemes  have been proposed theoretically in  Ref. [4-6] 
and realized experimentally in Ref. [8-10].

\par
Now that it can be made sure that some which-path detectors do not affect
the atomic center-of-mass wave function, according to Eqs. (1)-(2), the 
interference pattern should have  a contrast almost as good  as 
that when the which-path detector is not present.  Since the which-path
detector also tells us which-path each atom passes through, we now
see both the which-path information (particle-like) 
and the interference pattern (wave-like) can be obtained simultaneously. 
This,  of course, is in contradiction with the complementarity principle.
However, according to Ref. [4-10], the complementarity principle is still
upheld in these atom interferometers. Then, what happens to our simple
analysis? Or, what happens to the analysis made in Ref. [4-10]? One of
them must be wrong somewhere.
In the following, we will make a more detailed check on the theories and
experiments in Ref. [4-10].

\par
We  first come to the micromaser  which-path detector proposed  by 
Scully and coworkers [4-7], where two high-Q micromaser cavities
placed right before the double slits are used to detect the
which-path information of excited Rydberg atoms. They have 
shown [4] that no random phase fluctuation is caused by 
the electron-photon
interaction  to the center-of-mass wave function when the atom 
passes through the cavity.
 Subsequently they find that the total wave function for
the cavity photon field + atomic center-of-mass + electron system is
$$\Psi \propto (\psi_{1c} \psi_{1p}+\psi_{2c} \psi_{2p}) \psi_{e},
  \eqno(4)$$
where $\psi_{1p}=|1_1,0_2\rangle$ ($\psi_{2p}=|0_1,1_2\rangle$) 
denotes the field state when the atom passes through cavity 1 (2),
leaving a photon in cavity 1 (2), and deexciting the electron  into  lower
energy level $\psi_{e}$.
The interference pattern is simply  given by 
$P({\bf r})=  |\psi_{1c}({\bf r})|^2+|\psi_{2c}({\bf r})|^2$,
which is of null visibility. The reason is that the interference 
term $\psi_{1c}^{*}({\bf r})\psi_{2c}({\bf r})$ has a  coefficient
$\langle\psi_{1p}|\psi_{2p}\rangle=0$. Hence the complementarity principle
is enforced with the aid of entanglement
between the which-path detector and  the atomic motion.

\par
There are some problems in  above arguments.  Ever since the atom 
carrying the electron passes through the double slits, it immediately
ceases to interact with the which-path detector, because the photon in 
the cavity can not leak out of the cavities 
through the double slits. Then the 
wave function Eq. (4) is no longer applicable. In addition, 
it makes no sense to define the interference of atoms as the joint 
interference of the atom + electron + environment system.
The correct atomic wave function should be given by Eq. (1), 
the physical solution of Schr\"odinger's equation to diffraction 
of atomic wave by the double slits. Since there is no random phase
accumulated in the atomic center-of-mass motion as well as the 
electronic state during the passage through the which-path cavities,
 Eqs. (2) and (3) indicate that both waves will show visible interference 
pattern, in contradiction to the complementarity principle.   

\par
Since any theoretical assumption must be tested by experiments, 
let us turn to recent two realistic experiments on atom 
interferometers [8-10].  The  which-path detector
in the atom interferometer   designed by  Rempe and coworkers [8,9]
is  two internal electronic states of $^{85}$Rb, which can  
Rabi oscillate under  microwave pulses (for detail, see Ref. [8,9,13]). 
The  energy of microwave photon is far smaller  than the
$^{85}$Rb kinetic energy, so storage of the which-path information  
can only cause negligible change to the atomic center-of-mass motion.
The final interference pattern is observed by exciting the
atoms with a resonant laser beam and detecting the  fluorescence
photons.  The experimental results seem to  agree well with
the complementarity principle: When two microwave pulses are inserted,
the which-path information can be detected, and the interference
pattern is washed out. The authors of  Ref. [8,9] attribute their results 
 to the following double-slit type wave function:
$$\Psi \propto -\psi_{1c} \psi_{1e} +\psi_{2c} \psi_{2e}, \eqno(5)$$
where $\psi_{1c}$ ($\psi_{2c}$)  and  $\psi_{1e}$ ($\psi_{2e}$) are
the center-of-mass and electronic wave functions after the atom passes through 
path 1(2), respectively.  Since $\psi_{1e}$  and $\psi_{2e}$ are orthogonal,
no visible fringes appear in the curve $P=|\psi|^2$.   
Here the complementarity is also enforced through the entanglement between
the internal electronic state and the path taken by the atom, while the
uncertainty principle plays no role.

\par
The atom interferometer designed by Bertet {\it et al.} [10] uses a
Ramsey scheme, where a Rubidium Rydberg atom first at energy level $|e\rangle$ 
is split by the first microwave pulse $R_1$ 
into $|e\rangle$ and another  orthogonal lower state $|g\rangle$,
 then recombined by the second microwave pulse $R_2$
downstream, and finally detected by field-ionization (for detail, see
Ref. [10]). Here the microwave pulse does not influence the
atomic center-of-mass motion. The interference pattern is found
determined by the transition probability between levels $|e\rangle$ and
$|g\rangle$ [10], 
$$P_g(\phi)=\frac{1}{2}[1+{\rm Re}
(\langle\alpha_e|\alpha_g\rangle e^{i \phi})]. \eqno(6)$$
Here $\phi$ is the phase shift accumulated in the atomic center-of-mass
wave between two paths, $|\alpha_e\rangle$ and 
$|\alpha_g\rangle$ are the field states of $R_1$ microwave.
They are correlated with the electronic  states,  and  
 serve as the which-path detector. If the microwave field is originally
in vacuum state, then $|\alpha_e\rangle=|0\rangle$ and  
$|\alpha_g\rangle=|1\rangle$ are
orthogonal states (the which-path information has been recorded), 
Eq. (6) yields  invisible interference pattern. On the 
other hand, if the microwave field is originally in the 
classic state (coherent state) $|\alpha\rangle$, 
then $|\alpha_e\rangle$ and  $|\alpha_g\rangle$ are almost
identical to $|\alpha\rangle$ (no which-path information is recorded), 
and the interference pattern is recovered. 
Here once again  one finds perfect maintenance 
of the complementarity principle.

\par
Is it really the case that  above two experiments
 uphold the complementarity principle, as claimed by the authors?
The answer is no. On the contrary, they have implied the possibility
beyond the complementarity, if the atomic center-of-mass motion
(which, of course, is the mere motivation to design
 atom interferometers) is concerned. 
Above analysis clearly shows  that the interference pattern
in these two experiments is formed by the 
electron in the atom, instead of the atomic center-of-mass 
itself, since both  the fluorescence and the transition probability 
merely reflect the electronic states. One can also see this  
by comparing   Eqs. (5) and (6)  with Eq. (3). Obviously, 
they merely  reflect  the electronic motion.
In the first experiment, the wave function of the which-path detector
(electron) are prepared by the microwave pulses 
into form of Eq. (5), where the electronic states at the 
two slits are orthogonal to each other. No wonder
the interference  pattern completely disappears. 
In the second experiment, when the microwave field is in the vacuum state, 
the final electronic state in one path is 
  orthogonal to that in the other path, so the interference is destroyed.
 When the microwave field is in the classic state,
there are no visible difference in the electronic state 
in both paths, thus  the interference is recovered.
When the atomic center-of-mass motion is concerned, 
the interference pattern should keep visible in both experiments.
 Of course, in order to observe this interference pattern, one
must use  detectors that are only sensitive to the atomic center-of-mass
motion and behave  as if without seeing the electronic state.

\par
Now let us  say some words about the quantum eraser proposed in
Ref. [14,4-7,10]. In above analysis, we have shown that in
the micromaser interferometer, the final atomic wave function (either 
center-of-mass or electronic) after the atom has  passed through
the double slits, which determines the interference pattern, has
nothing to do with the micromaser field state. Therefore, erasing
the field state will cause nothing that can influence the
 interference pattern, not to say to determine the  pattern. 
In the experiment of
Ref. [10], the eraser, which is the second microwave pulse 
with an identical field state to the first pulse, 
is placed before the field-ionization  detector. 
The mere function of this eraser is to  compensate the difference
in the electronic state between the two paths caused by the first pulse, 
which  recovers the interference pattern. 
The reason why this good control
on the electronic state can be achieved is because the electron is bound
with the atom, so that it can only oscillate between two levels under
whatever complex influence by the microwave. For a free electron, this
task can not be accomplished.  The same holds for the atomic center-of-mass
motion. A quantum eraser directly exerting on the center-of-mass can not
completely delete the influence caused by the first detector, instead,
the uncertainty principle will dominate, and the interference pattern
can not be recovered. From this viewpoint,
 the quantum eraser proposed in Ref. [10] using a Mach-Zehnder's  
atom interferometer can not work at all.  The fact that 
 the which-path information can not be obtained does not necessarily
means  appearance of the interference pattern.  

\par
The above discussion on atom interferometers can be extended   to
other massive particles, such as neutrons. These particles have
internal freedoms like spins, which can serve as the which-path detector
in the interferometer. Numerous experiences on neutron interferometers
have shown that the change of spin state does not influence the 
coherence of neutron's  center-of-mass motion [15,16]. 
However,  to go beyond  the complementarity principle, 
one should first make sure 
that detectors only sensitive to the neutron's center-of-mass and
insensitive to the spins are available. How about  
massless particles like photons? In classic and
quantum physics, the vectorial nature (polarization) of 
photons (or electromagnetic waves) is well-established. Any interference
experiment of light must consider this vectorial nature [17].  
 It seems that  polarization is an inseparable part of photons.
As  polarization is the only  internal freedom
for photons, there is no which-path detector available for photons
in the same manner for massive particles
to go  beyond the complementarity principle.

\par
As a summary, we have revisited the fundamental 
complementarity principle and   wave-particle duality 
by studying atom interferometers in the framework of 
standard formulation of quantum mechanics.  Defining the 
interference of atom as  formed by the atomic center-of-mass
wave only, we  find it possible to use the internal freedoms of
 atoms, for example, the electronic state as a which-path detector
to obtain the which-path information of each atom, while
simulataneously conserve the coherence of the atomic 
center-of-mass motion  and  uphold
the interference pattern. In this manner, recent theoretical and 
experimental schemes of atom interferometers have
implied the possibility beyond the complementarity principle 
and refute of the wave-particle duality. 
Our theoretical analysis can be 
tested by experiments on atom interferometers similar to Ref. [8-10],
using  interference-pattern detectors 
that are directly sensitive of the atomic center-of-mass wave.

\end{document}